# A Calculator for Sediment Transport in Microchannels Based on the Rouse Number

L. Pekker

Fuji Film Dimatix Inc., Lebanon NH 03766 USA

**Abstract**

The Rouse number is commonly used to estimate the mode of the sediment transports in turbulent flows with large Reynolds number. However, in microchannels such as in modern inkjet systems, the liquid flows are usually laminar. In this paper, I modify the Rouse number by expanding it to the case of weakly turbulent and laminar flows and construct a calculator to estimate the modes of sediment transport in microchannels. To illustrate the applicability of the modified Rouse number, I apply it to the transport of sediments in an inkjet system and compare theoretical results with experimental observations. The modified Rouse number constructed in this paper can be used in other application as well.

## I. Introduction

In 1937, Hunter Rouse introduced a characteristic non-dimensional scale parameter [1], which later was named the Rouse number,

$$P = \frac{v_s}{u_* \kappa}, \tag{1}$$

that describes the modes of sediment transported in turbulent flows. In this equation, $v_s$ is the free fall settling (terminal) velocity of a sediment particle in the fluid, $\kappa = 0.4$ is the Karman constant calculated for turbulent flow, and $u_*$ is the boundary shear velocity determined as

$$u_* = \sqrt{\tau/\rho_f}, \tag{2}$$

where $\tau$ is the shear stress of the fluid at the bottom (at the sediment bed) and $\rho_f$ is the mass density of the fluid. Table 1 presents the transport modes of sediments vs. the Rouse number [2, 3]:

Table 1. Modes of sediment transport

| Mode of Transport | Rouse Number |
|---|---|
| Bed load | $P > 2.5$ |
| Suspended load: 50% Suspended | $1.2 < P < 2.5$ |
| Suspended load: 100% suspended | $0.8 < P < 1.2$ |
| Wash load | $P < 0.8$ |

Since, in turbulent flow, $u_*$ is proportional to the lift velocity of a particle at the sediment bed, Table 1 has perfect physical sense. Indeed, for large $P$, where the deposition rate of particles due to the gravity prevails over the particle lift, the sediments are transported as a bed load (in bed load mode); for small $P$, where the lift velocity of the particles is about the particle settling velocity, the particles are suspended in the flow and, therefore, the sediments are transported in the suspension mode; for very small $P$, where the particle lift velocity is much larger than the deposition rate of particles, the sediments are transported in the wash load mode. On the other hand, there should be a critical Rouse number that corresponds to a threshold for initiating the sediment motion; this is very important to know when designing microchannels to transport liquids with particles as, for an example, in the case of inkjet systems in which the ink consists of a liquid with micrometer-sized pigments. This threshold is described by the widely used the Shields diagram [4], which, unlike the Rouse number, is applicable for a large range of Reynolds numbers, including laminar flows as well, Fiq.1.

In Section II, I reformulate the Rouse number in Shields diagram terms using the particle boundary Reynolds number and the particle shear stress ($Re_*$ and $\tau_*$ in Fig.1) and then extend the Rouse number to weakly turbulent and laminar flows by matching the Rouse number to the Shields diagram. To illustrate the applicability of the modified Rouse number, I apply it to the transport of sediments in an inkjet system and compare theoretical results with experimental observations, Section III. Conclusions are given in Section IV.

## II. Expansion of the Rouse number to laminar flows

To connect the Rouse number with the Shields diagram, I will reformulate the Rouse number in terms of the Shields diagram. In the Shields diagram, Fig. 1, the particle boundary Reynolds number $Re_*$ and the particle shear stress $\tau_*$ are determined as

$$Re_* = \frac{D}{\mu}\sqrt{\tau \rho_f} \quad , \tag{3}$$

and

$$\tau_* = \frac{\tau}{Dg(\rho_p - \rho_f)}, \tag{4}$$

where $g$ is the Earth's gravitational acceleration; $D$ is the characteristic diameter of a particle; $\rho_p$ is the mass density of a particle; and $\mu$ is the fluid viscosity. In Eq. (1), I will use an approximate formula for the settling velocity of grains as given in [5],

$$v_s = \frac{gD^2(\rho_p - \rho_f)}{C_1\mu + \sqrt{0.75C_2(\rho_p - \rho_f)\rho_f g D^3}} \tag{5}$$

with coefficients $C_1$ and $C_2$ from Table 2 [5]:

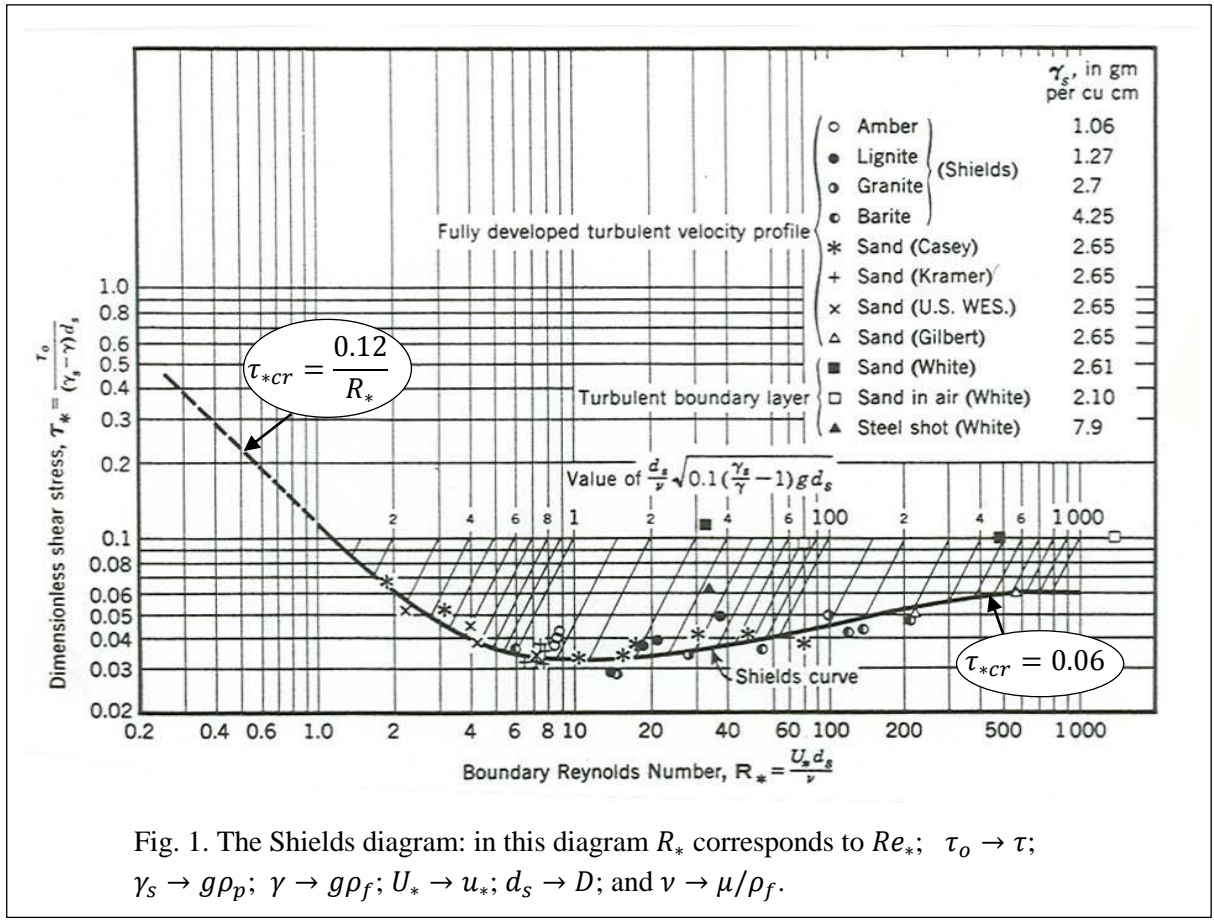

Fig. 1. The Shields diagram: in this diagram $R_*$ corresponds to $Re_*$; $\tau_o \to \tau$; $\gamma_s \to g\rho_p$; $\gamma \to g\rho_f$; $U_* \to u_*$; $d_s \to D$; and $\nu \to \mu/\rho_f$.

Table 2. Coefficients for settling velocity of grains

| Constant | Smooth Sphere | Natural Grains: Sieve Diameters | Natural Grains: Nominal Diameters | Limit for Ultra-Angular Grains |
| --- | --- | --- | --- | --- |
| $C_1$ | 18 | 18 | 20 | 24 |
| $C_2$ | 0.4 | 1.0 | 1.1 | 1.2 |

As one can see from Eq. (5), in the creeping free fall case where the first term in the dominator of Eq. (5) is much larger than the second term, Eq. (5) reduces to the Stokes formula for settling velocity,

$$v_s = gD^2(\rho_p - \rho_f)/(C_1\mu), \tag{6}$$

and, in the turbulent free fall case where the first term is much smaller than the second term, Eq. (5) reduces to the formula for the turbulent settling velocity,

$$v_s = \sqrt{\frac{4gD}{3C_2}\left(\frac{\rho_p - \rho_f}{\rho_f}\right)} \quad , \tag{7}$$

where $C_1$ and $C_2$ are the creepy and turbulent drag coefficients respectively.

Substituting Eqs. (2) and (5) into Eq. (1) and then using Eqs. (3) and (4), Eq. (1) can be reduced to the following form:

$$P = \frac{1}{\kappa C_1}\left(\frac{Re_*}{\tau_*}\right)\left(1 + \sqrt{\frac{0.75 C_2}{(C_1)^2}}\left(\frac{Re_*}{\sqrt{\tau_*}}\right)\right)^{-1} . \tag{8}$$

Asymptotic solutions of Eq. (8) are:

$$\tau_*(Re_* \to \infty) = \frac{4}{3C_2(\kappa P)^2} \quad , \tag{9}$$

$$\tau_*(Re_* \to 0) = \frac{Re_*}{\kappa C_1 P} . \tag{10}$$

Solving Eq. (8) for $\tau_*$, I obtain that $\tau_*$ can be expressed in terms of $Re_*$ and $P$ as follows:

$$\tau_* = \left(-\left(\frac{0.75 \cdot C_2 \cdot Re_*^2}{4 C_1^2}\right)^{0.5} + \left(\frac{0.75 \cdot C_2 \cdot Re_*^2}{4 C_1^2} + \frac{Re_*}{\kappa C_1 P}\right)^{0.5}\right)^2 \tag{11}$$

The Shields diagram also has two asymptotes [1]:

$$\tau_{*cr}(Re_* \to \infty) = 0.06 , \tag{12}$$

$$\tau_{*cr}(Re_* \to 0) = \frac{0.12}{R_*} . \tag{13}$$

To match the Rouse number and the Shields diagram, I introduce the critical value of the Rouse number, $P_{cr}$, that yields the same $\tau_*$ as the Shields diagram at $Re_* \gg 1$ and corresponds to the threshold of initiation of the sediment motion at the bed. Setting Eqs. (9) and (12) equal to each other, I obtain

$$P_{cr} = \frac{11.79}{\sqrt{C_2}} . \tag{14}$$

In Eq. (14), I have taken into account that the Karman constant is equal to 0.4.

Fig. 2 shows the Shields curve, $\tau_{*cr}(Re_*)$, and curves of $\tau_*(Re_*)$ calculated by Eq. (7) for different modes of sediment transport in the case of ultra-angular grains; $C_1$ and $C_2$ are taken from Table 2. As one can see from Fig. 2, in the case of the threshold of initiation of sediment motion mode, $\tau_*(P_{cr}, Re_*)$ is in a good agreement with the Shields diagram when $Re_* > 10$; however, with a decrease in $Re_*$, $\tau_*(P_{cr}, Re_*)$ sharply diverges from the Shields diagram. For other sediment transport modes, $\tau_*$ has no sense for small $Re_*$ as well. This is so because Eq. (1) is applicable for turbulent flows only, for large $Re_*$.

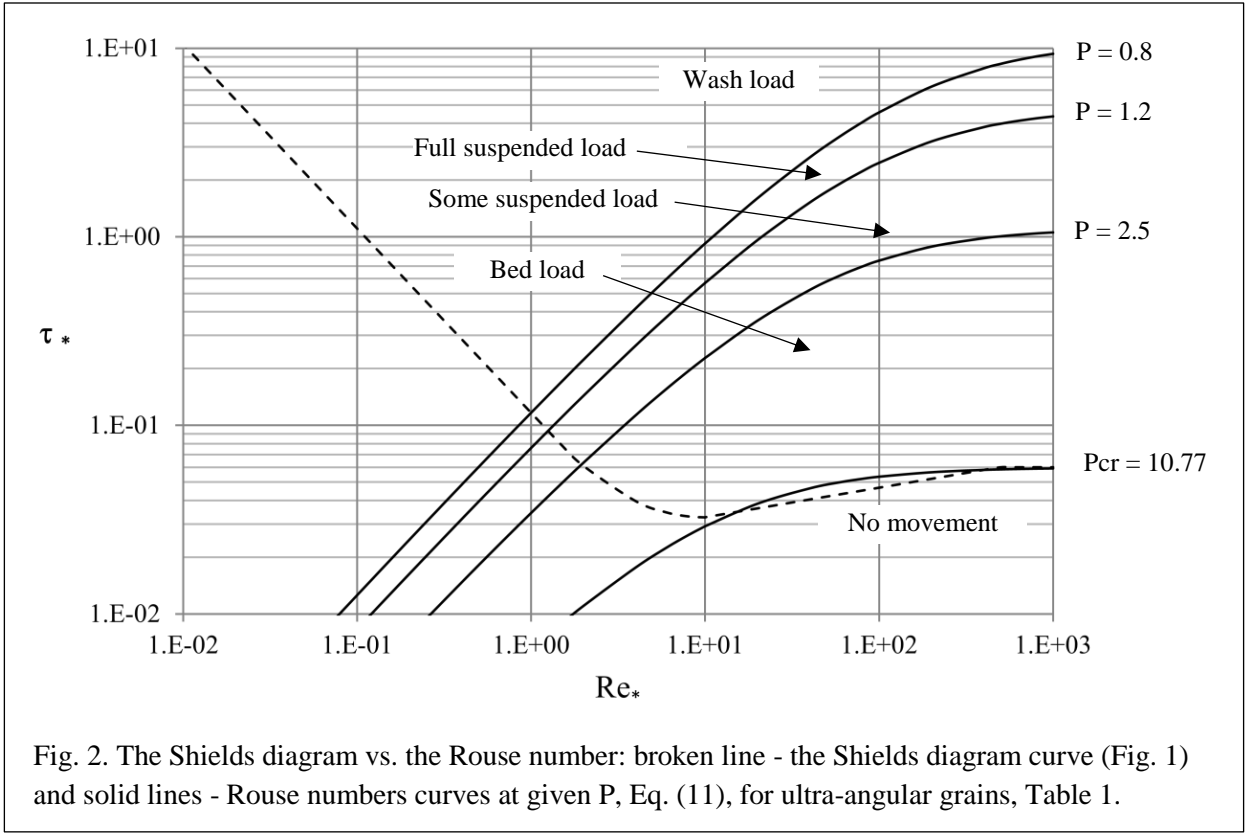

Fig. 2. The Shields diagram vs. the Rouse number: broken line - the Shields diagram curve (Fig. 1) and solid lines - Rouse numbers curves at given P, Eq. (11), for ultra-angular grains, Table 1.

To expand the Rouse number to small $Re_*$ and, at the same time to preserve the asymptote of $\tau_*(P_{cr}, Re_* \to \infty)$, Eq. (9) with $P = P_{cr}$, I will modify the Rouse number by using $\tilde{\kappa}$,

$$\frac{1}{\tilde{\kappa}} = \frac{1}{\kappa} + \frac{0.12 C_1 P_{cr}}{(Re_*)^2}, \tag{15}$$

for $\kappa$ in Eq. (11); this yields

$$\tau_* = \left( -\left(\frac{0.75 \cdot C_2 \cdot Re_*^2}{4C_1^2}\right)^{0.5} + \left(\frac{0.75 \cdot C_2 \cdot Re_*^2}{4C_1^2} + \frac{Re_*}{0.4 C_1 P} + \frac{0.12 P_{cr}}{P Re_*}\right)^{0.5} \right)^2. \qquad (16)$$

The substitution of $\tilde{\kappa}$ for $\kappa$ into Eqs. (9) and (10) shows that the asymptotes of Eqs. (16) for $Re_* \gg 1$ and $Re_* \ll 1$ indeed coincide with Eqs. (12) and (13) respectively. Fig. 3 demonstrates the excellent agreement between the Shields diagram curve and $\tau_*(P_{cr}, Re_*)$.

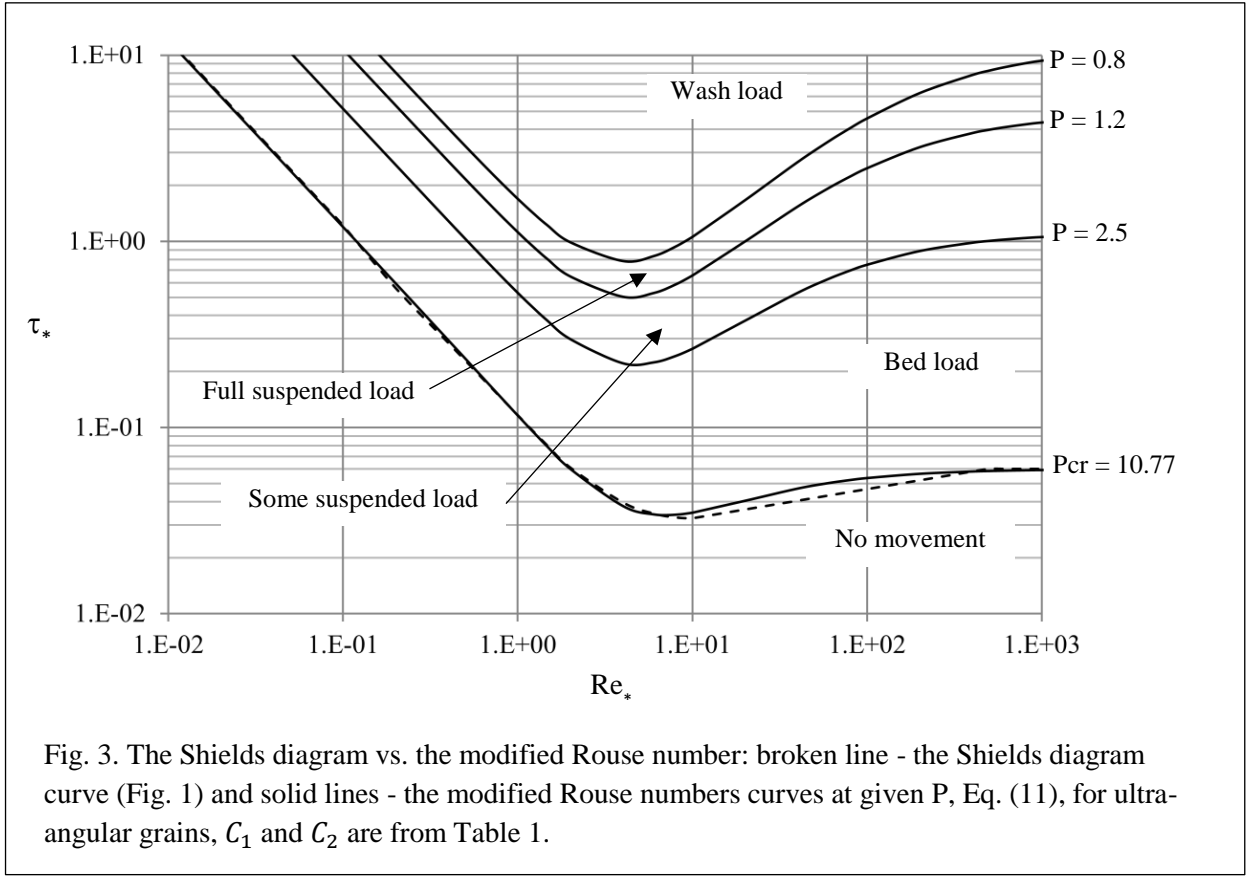

Fig. 3. The Shields diagram vs. the modified Rouse number: broken line - the Shields diagram curve (Fig. 1) and solid lines - the modified Rouse numbers curves at given P, Eq. (11), for ultra-angular grains, $C_1$ and $C_2$ are from Table 1.

## III. Comparison with experiment

To illustrate the applicability of the modified Rouse number, I applied it to the transport of sediments in a channel of Fuji Dimatix SG 1024 inkjet head and compared theoretical results with experimental observations. I used Microsoft Excel to plot $\tau_*$ vs. $Re_*$ for the sediment transport modes from Table 1 using Eq. (16) and to plot the threshold of initiation of the sediment motion at the bed mode that corresponds to $P = P_{cr}$ calculated by Eq. (14). These plots are shown in the "plot window" of the calculator along with the Shields diagram curve, $\tau_{*cr}(Re_*)$. The calculator input parameters are the characteristic radius of particles, $C_1$ and $C_2$; the density and viscosity of the fluid; the shear stress at the sediment bed or both the radius of the cylindrical channel and the liquid flow. In the case where input parameters are the flow and the diameter of the channel, I calculate the shear stress at the bottom assuming Poiseuille's law, which is reasonable for this particular application.

Table 3 presents the combined Excel input and output data for a channel of Fujifilm Dimatix SG 1024 inkjet print head for two ink flows: $2.71 \cdot 10^{-8}$ and $5.646 \cdot 10^{-7}$ m$^3$/sec. The combined "screen shots" of the calculator plot window for these flows are presented in Fig. 4. As one can see from this figure, in the case of small flow, the sediment transport is in no movement mode; while in the case of large flow, it is in wash load mode. The experimental work with this type of print heads showed that, for the ink flow of $2.71 \cdot 10^{-8}$ m$^3$/sec, all nozzles of the print head where completely blocked by pigment particles and the print head could not be recovered; this experimental result corresponds to the case of "small flow" in Fig. 4. However, for the ink flow of $5.646 \cdot 10^{-7}$ m$^3$/sec, the print head worked normally and the particles were transported through this channel; this regime corresponds to the case of "large flow" in Fig. 4. Thus, experimental findings support the theoretical model

Table 3

| **Input Parameters** | **Small Flow** | **Large Flow** |
|---|---|---|
| Particle Diameter, D (m) | 5.0E-06 | 5.0E-06 |
| Particle Density, $\rho_p$ (kg/m$^3$) | 1320 | 1320 |
| Fluid Density, $\rho_w$ (kg/m$^3$) | 1000 | 1000 |
| Viscosity, (Pa*s) | 0.015 | 0.015 |
| C1 | 24 | 24 |
| C2 | 1.2 | 1.2 |
| Flow, Q (m$^3$/s) | 2.71E-8 | 5.646E-7 |
| Radius of the pipe, R (m) | 1.5E-03 | 1.5E-03 |
| **Output Parameters** | | |
| Shear Stress, $\tau$ (Pa) | 0.153 | 3.194 |
| Boundary Reynolds Number, R$_*$ | 0.004 | 0.019 |
| Boundary shear stress, $\tau_*$ | 9.770 | 203.518 |
| Pcr | 10.76 | 10.76 |
| Shield's Shear Stress, $\tau_{*P=Pcr}$ | 29.070 | 6.368 |
| Shear Stress Some Suspension, $\tau_{*P=2.5}$ | 125.239 | 27.438 |
| Shear Stress Full Suspension, $\tau_{*P=1.2}$ | 260.916 | 57.165 |
| Shear Stress Wash Load, $\tau_{*P=0.8}$ | 391.375 | 85.748 |
| Resuspension? | **No Movement** | **Wash Load** |

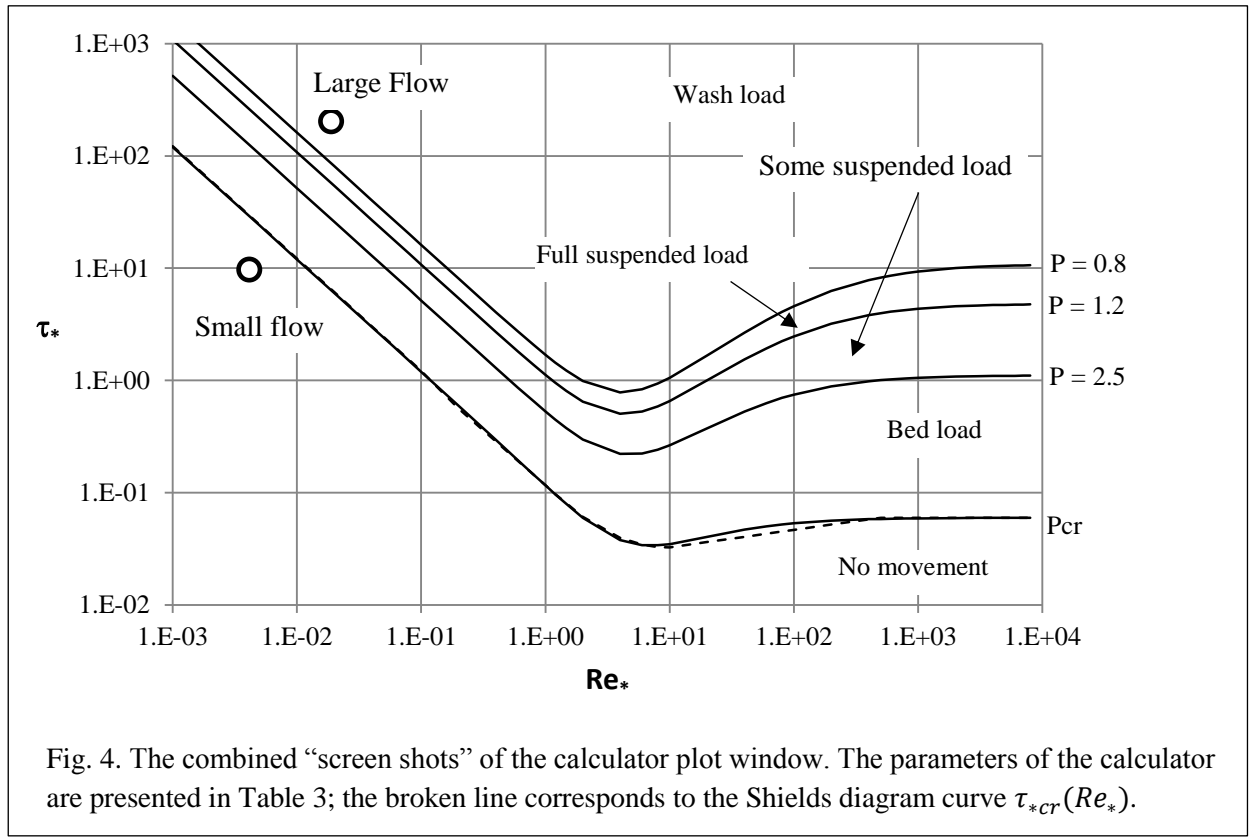

Fig. 4. The combined "screen shots" of the calculator plot window. The parameters of the calculator are presented in Table 3; the broken line corresponds to the Shields diagram curve $\tau_{*cr}(Re_*)$.

## IV. Conclusions

In this paper, I modified the Rouse number by expanding it to the case of weakly turbulent and laminar flows by matching it to the Shields diagram curve for the threshold of initiation of the sediment motion mode and demonstrating excellent agreement between the Shields diagram and the Rouse number calculated for this transport mode. Based on the modified Rouse number model, I have constructed an Excel calculator to estimate the transport of sediments in microchannels and applied it a Fujifilm Dimatix print head, finding strong correlation between theoretical results and experimental observations. The modified Rouse number constructed in this paper can be used in other applications as well, for example in hydrology.

**Acknowledgments**

I would like to express my sincere gratitude to my colleagues Daniel Barnett and Matthew Aubrey for their kind support and helpful discussions during this research. I also would like to thank Alexander Pekker for his kind help in preparation the text of this paper.
**References**

[1] Rouse H., *Modern Conceptions of Mechanics of Fluid Turbulence*, Trans. ASCE, Vol. 102, No. 1, pp. 463-505 (1937), http://cedb.asce.org/CEDBsearch/record.jsp?dockey=0288088.

[2] Whipple, Kelin (September 2004). "IV. Essentials of Sediment Transport"(PDF). *12.163/12.463 Surface Processes and Landscape Evolution: Course Notes*. MIT OpenCourseWare, Retrieved 2009-10-11.

[3] Moore, Andrew. "Lecture 20—Some Loose Ends" (PDF). *Lecture Notes: Fluvial Sediment Transport*, Kent State, Retrieved 23 December 2009.

[4] Shields A., *Anwedung der Aehnlichkeysmechanik und der Turbulenzforsschung auf die Geschiebebewegung. Mitteilungen der Pruessischen Versuchanstalt fur Wasserbau und Schiffbau*, Berlin 1936 (uuid:61a19716-a994-4942-9906-f680eb9952d6); see also in Garde R.J and Ranga Raju K.G., *Mechanics of Sediment Transportation and Alluvial Stream Problems*, Third Edition, New Age International Publishers (2000).

[5] Ferguson, R. I., and M. Church (2006), *A Simple Universal Equation for Grain Settling Velocity, Journal of Sedimentary Research*, 74(6) 933-937, doi:10.1306/051204740933.
10